\documentclass[times,astrobib,amssymb]{mn2e}
\usepackage{psfig}

\newcommand{\be}{\begin{equation}}
\newcommand{\en}{\end{equation}}

\def\zabs{$z_{\rm abs}$}
\def\zem{$z_{\rm em}$~}

\def\mgii{Mg~{\sc ii}~}

\def\kms{km~s$^{-1}$}
\title[Dusty 21-cm absorbers at z$\sim$1.3]{Detection of the 2175~\AA~extinction feature and 
21-cm absorption in two Mg~{\sc ii} systems at $z\sim1.3$}
\author[Srianand et al.]{R. Srianand$^{1}$\thanks{E-mail: anand@iucaa.ernet.in}, N. Gupta$^2$, P. Petitjean$^3$, P. Noterdaeme$^3$ and
D. J. Saikia$^4$\\
$^{1}$ IUCAA, Post Bag 4,  Ganeshkhind, Pune 411 007, India \\
$^{2}$ ATNF, CSIRO, P. O. Box 76, Epping, NSW 1710, Australia\\
$^{3}$ UPMC Paris 6, Institut d'Astrophysique de Paris, UMR7095 CNRS, 98bis Boulevard Arago, 75014, Paris, France\\
$^{4}$ NCRA, Post Bag 3,  Ganeshkhind, Pune 411 007, India }

\begin{document}

\date{Accepted. Received; in original form }

\pagerange{\pageref{firstpage}--\pageref{lastpage}} \pubyear{2005}

\maketitle

\label{firstpage}

%%%%%%
\begin{abstract}

We have discovered two dusty intervening Mg~{\sc ii} absorption systems at $z\sim1.3$ in the Sloan Digital Sky Survey (SDSS) database. 
The overall spectra of both QSOs are red
(u-K$>$4.5 mag) and are well modelled by the composite QSO 
spectrum reddened by the extinction curve from the Large Magellanic 
Cloud(LMC2) Supershell redshifted to the rest-frame of the Mg~{\sc ii} 
systems. In particular, we detect clearly the presence of the UV 
extinction bump at $\lambda_{\rm rest}\sim 2175$~\AA.
Absorption lines of weak transitions like 
Si~{\sc ii}$\lambda$1808, Cr~{\sc ii}$\lambda$2056, 
Cr~{\sc ii}+Zn~{\sc ii}$\lambda$2062, 
Mn~{\sc ii}$\lambda$2594, Ca~{\sc ii}$\lambda$3934 
and Ti~{\sc ii}$\lambda$1910 from these systems 
are detected even in the low signal-to-noise ratio and low 
resolution SDSS spectra, suggesting high column densities
of these species. The depletion pattern 
inferred from these absorption lines is consistent
with that seen in the cold neutral medium of the LMC. 
Using the LMC $A_V$ vs. $N$(H~{\sc i}) relationship we derive 
$N$(H~{\sc i})$\sim 6\times 10^{21}$ cm$^{-2}$ in both systems. 
Metallicities are close to solar.
Giant Metrewave Radio Telescope (GMRT) observations of 
these two relatively weak radio loud QSOs (f$_\nu$ $\sim$ 50 mJy) resulted
in the detection of 21-cm absorption in both cases. 
We show that the spin temperature of the gas is
of the order of or smaller than 
$500$~K.
These systems provide a unique opportunity to search for molecules
and diffuse interstellar bands at $z>1$.
\end{abstract}
 
%%%%%%%%%%%%%%%%%%%%%%%%%%%%

\section{Introduction}

Studying the physical conditions in the interstellar medium (ISM) 
at high redshift and the processes that
maintain these conditions is important for our understanding of how
galaxies form and evolve.
The presence of dust influences the physical state of the gas through 
photo-electric heating,  UV shielding, and formation of molecules on 
the surface of grains. However, we know very little about 
dust properties in the ISM at high redshifts. 
Properties of the dust can be derived from extinction curves
observed in different astrophysical objects. Recently, Noll et al. (2007)
found evidence for the presence of a moderate UV bump 
in at least part of the population of massive galaxies at $z>1$.
Studies of dust extinction in the circum-burst environment of Gamma
Ray Bursts indicate that in most cases, a LMC-like extinction curve
is preferred (e.g. Heng et al. 2008).

The depletion of Cr with respect to Zn in intervening damped Lyman-$\alpha$ 
systems (DLAs)  shows that dust is indeed 
an important component of the high density gas (Pettini, Smith \& Hunstead 1994).
A correlation is observed in DLAs between metallicity and dust-depletion  
(Ledoux, Petitjean \& Srianand 2003) which
is confirmed by the higher detection rate of H$_2$ in DLAs 
with higher metallicities (Petitjean et al. 2006; Noterdaeme et al. 2008).
The corresponding gas is cold (T$\sim$150 K; Srianand et al. 2005) 
as expected from multiphase ISM models 
in which the gas with high metallicity and dust content 
has lower kinetic temperature than the gas with 
lower metallicity and dust content (Wolfire et al. 1995). 
However, even in the highest metallicity DLAs 
typical dust signatures like high extinction (i.e 
0.16$\le$ E(B$-$V)$\le$0.40), 2175~\AA~absorption bump 
or the diffuse interstellar bands (DIBs) are
not seen.

Signatures of dust are seen from few intermediate
and low redshift absorption systems. 
Both DIBs and 2175\AA~absorption bump
have been detected in the $z_{\rm abs}$~=~0.524 
system toward AO~0235+164
(Wolfe \& Wills 1977; York et al. 2006;
Junkkarinen et al. 2004; Kulkarni et al. 2007).
This system also shows strong 21-cm 
absorption (Wolfe \& Wills 1977)
and has $N$(H~{\sc i}) = 5$\times10^{21}$ cm$^{-2}$,
E(B-V) = 0.23 (Junkkarinen et al. 2004) and 
$T_{\rm s}$ = 220$\pm$60 K.
Wang et al. (2004) have 
reported the detection of the 2175\AA~absorption bump in 3
intermediate redshift ($z\sim1.3$) Mg~{\sc ii} 
systems. 
 Recently, Ellison et al. (2008) have detected
DIBs in the \zabs = 0.1556 Ca~{\sc ii} systems towards SDSS J001342.44-002412.6.
Composite spectra have been obtained for different samples of 
absorbers from the Sloan Digital Sky Survey (SDSS). They show in a 
statistical way that dust 
is present in strong Mg~{\sc ii} (York et al. 2006a)  and Ca~{\sc ii} 
systems (Wild, Hewett \& Pettini 2006). 
In the former case
the mean extinction curve is similar to the 
SMC curve with a rising 
ultraviolet extinction below 2175~\AA~ with E(B$-$V)$\le$0.08 
and with no evidence of an UV bump. In the latter case, 
evidence for the UV bump is marginal and LMC extinction
curve provides E(B-V)$\le$0.103 for different sub-samples.

An efficient way to reveal cold and dusty gas is to search for 21-cm absorption 
as shown by the high detection rate of 21-cm absorption towards red QSOs
(Carilli et al. 1998; Ishwara-Chandra, Dwarakanath \& Anantharamaiah 2003; 
Curran et al. 2006). 
Multiple lines of sight toward lensed QSOs passing 
through regions producing high extinction also show 21-cm and molecular absorption lines 
(see Wiklind \& Combes 1996). 

We have recently completed a systematic search for 
21-cm absorption in a complete sample of
38 Mg~{\sc ii} systems, drawn from the SDSS DR5,
with redshifts in the range $1.10\le z\le 1.45$ 
corresponding to the frequency coverage of
610~MHz feed at GMRT (see Gupta et al. 2007 for
early results). 
Using an automatic procedure we have
identified 5 systems in this sample 
showing strong absorption lines from 
Si~{\sc ii}, Zn~{\sc ii}, Cr~{\sc ii}, 
Fe~{\sc ii}, Mg~{\sc ii} and Mg~{\sc i} in front
of radio-loud QSOs with flux density $\ge50$ mJy. 
Here, we report the detection of 21-cm absorption from two
of these systems at \zabs$\sim$1.3 towards
red (u-K$\ge$4.5 mag) QSOs. The optical spectra of
these two quasars possess 2175\AA~dust absorption features. 

\section{Observations and analysis:}

\begin{table}
\caption{GMRT observation log and results}
\begin{center}
\begin{tabular}{l c c c c c}
\hline
\hline
Source name  & Date          & Time   &Peak Flux$^a$ & rms$^b$ \\
             &               & (hr)   &  &  \\
\hline
J0850+5159  &   2007 Nov 06 &7.3     &   64.0 &  1.2  \\
J0852+3435  &   2007 Nov 05 &7.5     &   51.2 &  2.4   \\
                      &   2007 Nov 30 &6.9     & 52.2&1.3        \\
                      &   2008 Mar 08 &6.2     & 50.1&1.3        \\
\hline
\end{tabular}
\end{center}
\begin{flushleft}
$^a$ in units of mJY beam$^{-1}$ \\
$^b$ spectral rms in units of mJy\,beam$^{-1}$\,channel$^{-1}$.
\end{flushleft}
\label{gmrtlog}
\end{table}

\begin{figure}
\centerline{{
\psfig{figure=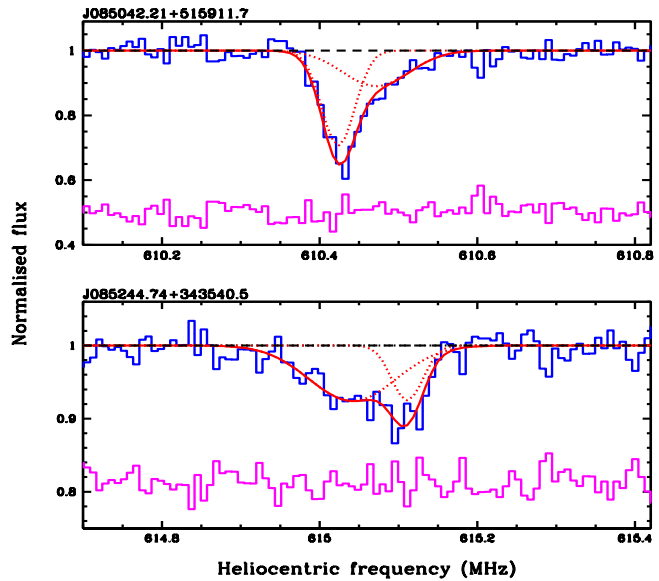,height=8.0cm,width=9.0cm,angle=0}
}}
\caption[]
{GMRT spectra of J085042.21+515911.7 (top panel) and
J085244.74+343540.5 (bottom panel). H~{\sc i} 21-cm absorption
is detected at $z_{\rm abs}$~=~1.3265 and 1.3095 (see Table~2),
respectively. 
The solid lines represent the fits to the overall profiles.
Individual Gaussian components are shown with dotted lines.
Residuals, on a scale arbitrarily shifted
       for clarity, are also plotted.
}
\label{gmrtdet}
\end{figure}

\begin{table}
\caption{Details of multiple Gaussian fits to 21-cm absorption lines.}
\begin{center}
\begin{tabular}{l c c c c}
\hline
\hline
QSO & z$_{\rm abs}$ & $\Delta v^{\rm a}$ & $\tau_{\rm p}$ &  {$N$(H~I) $f_{\rm c}^b \over T_{\rm s}$}\\
\hline
J0850+5159 &1.32674  &49$\pm$16   &0.117$\pm$0.036  &1.11$\pm$0.50 \\ 
           &1.32692  &24$\pm$4    &0.347$\pm$0.108  &1.61$\pm$0.57 \\ 
J0852+3435 &1.30919  &23$\pm$7    &0.078$\pm$0.025  &0.35$\pm$0.15 \\ 
           &1.30945  &63$\pm$12   &0.079$\pm$0.009  &0.96$\pm$0.21 \\ 
\hline
\end{tabular}
\end{center}
\begin{flushleft}
$^a$ FWHM in km s$^{-1}$; $^b$ in units of 10$^{19}$ cm$^{-2}$ K$^{-1}$
\end{flushleft}
\label{gmrtres}
\end{table}

We observed J085042.21+515911.7 (called J0850+5159) and 
J085244.74+343540.5 (called J0852+3435) with the GMRT 
to search for 21-cm absorption associated with the 
\mgii systems at \zabs = 1.3265 and \zabs = 1.3095 respectively. 
Log of our observations is presented in Table~\ref{gmrtlog}.  
We used 1 MHz baseband bandwidth split into 128 frequency channels 
(corresponding to a velocity resolution of $\sim$3.8 \kms) centered around 
the redshifted 21-cm frequency. 
We observed standard flux density calibrators 3C\,147 and 3C\,286 every 
2-3 hours to correct for amplitude and bandpass variations.  Phase 
calibrators (J0834+555 for J0850+5159 and J0741+312 for 
J0852+3435) were also observed approximately every 40 min.  
Data were acquired in both the 
circular polarization channels, RR and LL
%These data were 
and reduced in the standard way using the Astronomical Image Processing 
System (AIPS) as described in Gupta et al. (2006). 
GMRT spectra of both sources show detectable 
21-cm absorption (see Fig.~\ref{gmrtdet}).
The broad absorption produced by the
\zabs = 1.3095 system towards SDSS J0852+3435 is confirmed by 
repeated GMRT observations obtained at three 
different epochs spread over 4 months. We have used
the combined spectrum for the analysis presented here. 
It is usual procedure to decompose the absorption profile into
multiple Gaussian components. For a Gaussian profile the 
H~{\sc i} column density and the peak optical depth ($\tau_{\rm p}$) 
are related by,
\begin{equation}
N({\rm HI})f_{\rm c} = 1.93 \times 10^{18}~\tau_{\rm p}~T_{\rm s}~\Delta v~{\rm cm}^{-2}.
\end{equation}
Here, $\Delta v$ and $f_{\rm c}$ are the FWHM (in \kms) of the fitted Gaussian component, 
and the covering factor of the absorbing cloud respectively. Results of the Gaussian fits to the 21-cm absorption lines are given in Table~\ref{gmrtres}. 

Optical spectra of the QSOs were downloaded from the SDSS archive. 
\begin{table}
\caption{Rest equivalent widths of UV absorption lines.}
\begin{tabular}{l c c c}
\hline
Species & \multicolumn{3}{c}{Rest equivalent widths (\AA)}\\
        & J0850$+$5159 & J0852$+$3436 & \multicolumn{1}{c}{Ca~{\sc ii}}\\
        & \zabs=1.3265 & \zabs=1.3095 &composite\\
\hline
\hline
Ca~{\sc ii}$\lambda$3934 &.... &1.05&0.49$-$1.01\\
Si~{\sc ii}$\lambda$1808 &0.96 &1.06& ....\\
Mg~{\sc ii}$\lambda$2798 &4.62 &2.89&2.19$-$2.47\\
Mg~{\sc ii}$\lambda$2803 &4.15 &2.80&2.07$-$2.23\\
Mg~{\sc i}$\lambda$2853  &2.08 &1.11&0.72$-$0.93\\
Mn~{\sc ii}$\lambda$2576 &0.91 &1.25&0.20$-$0.40\\
Mn~{\sc ii}$\lambda$2594 &0.43 &1.30&0.15$-$0.23\\
Mn~{\sc ii}$\lambda$2606 &0.50 &0.63&0.10$-$0.12\\
Fe~{\sc ii}$\lambda$2249 &0.58 &$<0.56$&0.09$-$0.10\\
Fe~{\sc ii}$\lambda$2260 &0.55 &$<0.56$&0.08$-$0.08\\ 
Fe~{\sc ii}$\lambda$2344 &1.61 &1.46&1.36$-$1.39\\ 
Fe~{\sc ii}$\lambda$2374 &1.72 &1.47&0.89$-$1.01\\
Fe~{\sc ii}$\lambda$2382 &2.19 &2.36&1.55$-$1.61\\ 
Fe~{\sc ii}$\lambda$2586 &1.69 &1.60&1.27$-$1.35\\ 
Fe~{\sc ii}$\lambda$2600 &2.27 &2.10&1.60$-$1.80\\
Zn~{\sc ii}+Mg~{\sc i}$\lambda2026$ &0.98&0.85&0.12$-$0.27\\
Cr~{\sc ii}$\lambda$2056 &0.67 &$<0.30$&0.09$-$0.11\\
Cr~{\sc ii}+Zn~{\sc ii}$\lambda2062$ &1.12&1.20&0.11$-$0.20\\
Cr~{\sc ii}$\lambda$2066 &0.50 &0.39&0.08$-$0.08\\
Ti~{\sc ii}$\lambda$1910 &$<0.10$&$<0.20$& ....\\
Ti~{\sc ii}$\lambda$3242 &0.33 &0.60&0.09$-$0.12\\
Ti~{\sc ii}$\lambda$3385 &0.63 &0.71&0.09\\
\hline
\end{tabular}
\label{sdssuv}
\end{table}
\begin{table}
\caption{Results of SED fitting to the SDSS spectrum.}
%\begin{center}
\begin{tabular}{ccclc}
\hline
Object & \zem &Dust & \multicolumn{1}{c}{$A_{V_a}$} & $\chi_\nu^2$\\
       & & model&           &             \\
\hline
\hline
J0850$+$5159&1.89&MW   &0.83(0.06)$^1$ & 2.75\\
                &&SMC  &0.59(0.05)$^1$ & 1.42\\
                &&     &0.51(0.04)$^2$ & 1.60\\
                &&     &0.51(0.04)$^3$ & 1.36\\
                &&LMC2 &0.83(0.06)$^1$ & 1.16\\
                &&     &0.73(0.05)$^2$ & 1.43\\
                &&     &0.70(0.05)$^3$ & 1.27\\
J0852$+$3435&1.65&MW   &1.20(0.08)$^1$ & 2.49\\
                &&SMC  &0.75(0.06)$^1$ & 2.09\\
                &&     &0.64(0.05)$^2$ & 2.14\\
                &&     &0.68(0.06)$^3$ & 1.99 \\ 
                &&LMC2 &1.10(0.08)$^1$ & 1.42\\
                &&     &0.95(0.08)$^2$ & 1.64\\
                &&     &0.97(0.08)$^3$ & 1.53\\ 
\hline
\end{tabular}
%\end{center}
\begin{flushleft}
$^1$ using the SDSS QSO composite spectrum (Vanden Berk et al. 2001); $^2$ using the HST QSO composite spectrum (Zheng et al. 1997);
$^3$ using LBQS QSO composite spectrum (Francis et al. 1991).
\end{flushleft}
\label{tchi}
\end{table} 
In Table~\ref{sdssuv} we summarize the rest equivalent widths of
various metal absorption lines in the two  absorption 
systems and compare them
with the range of equivalent widths observed for the Ca~{\sc ii}
systems by Wild, Hewett \& Pettini (2006). A glance at this 
table suggests that the  two systems discussed here 
have much stronger absorption lines than 
those associated with the Ca~{\sc ii} systems. 

Next we compute the extinction due to the Mg~{\sc ii} 
system by assuming the reddening of the quasar to be a 
consequence of the presence of dust at \zabs.
The optical depth at an observed wavelength $\lambda$ 
can be written as,
\begin{equation}
 \tau(\lambda) = 0.92 {A_{\lambda_{\rm a}}} = 0.92 A_{V_a} \xi (\lambda_a).
\end{equation}
Here, ${A_{\lambda_{\rm a}}}$ and ${A_{V_{a}}}$ are 
the extinction in magnitude at $\lambda_{\rm a}$~=~$\lambda/(1+z_{\rm abs})$ and at the rest V band of the absorber
respectively. $\xi(\lambda_a)$ is the relative extinction 
at ${{\lambda_{\rm a}}}$ to that in the rest V band.
We consider $\xi(\lambda)$ for the SMC, the LMC2 Supershell and the Galaxy
(Misselt, Clayton \& Gordon 1999; Gordon et al. 2003) 
in order of increasing UV bump strength.

\begin{figure}
%\centerline{{
\psfig{figure=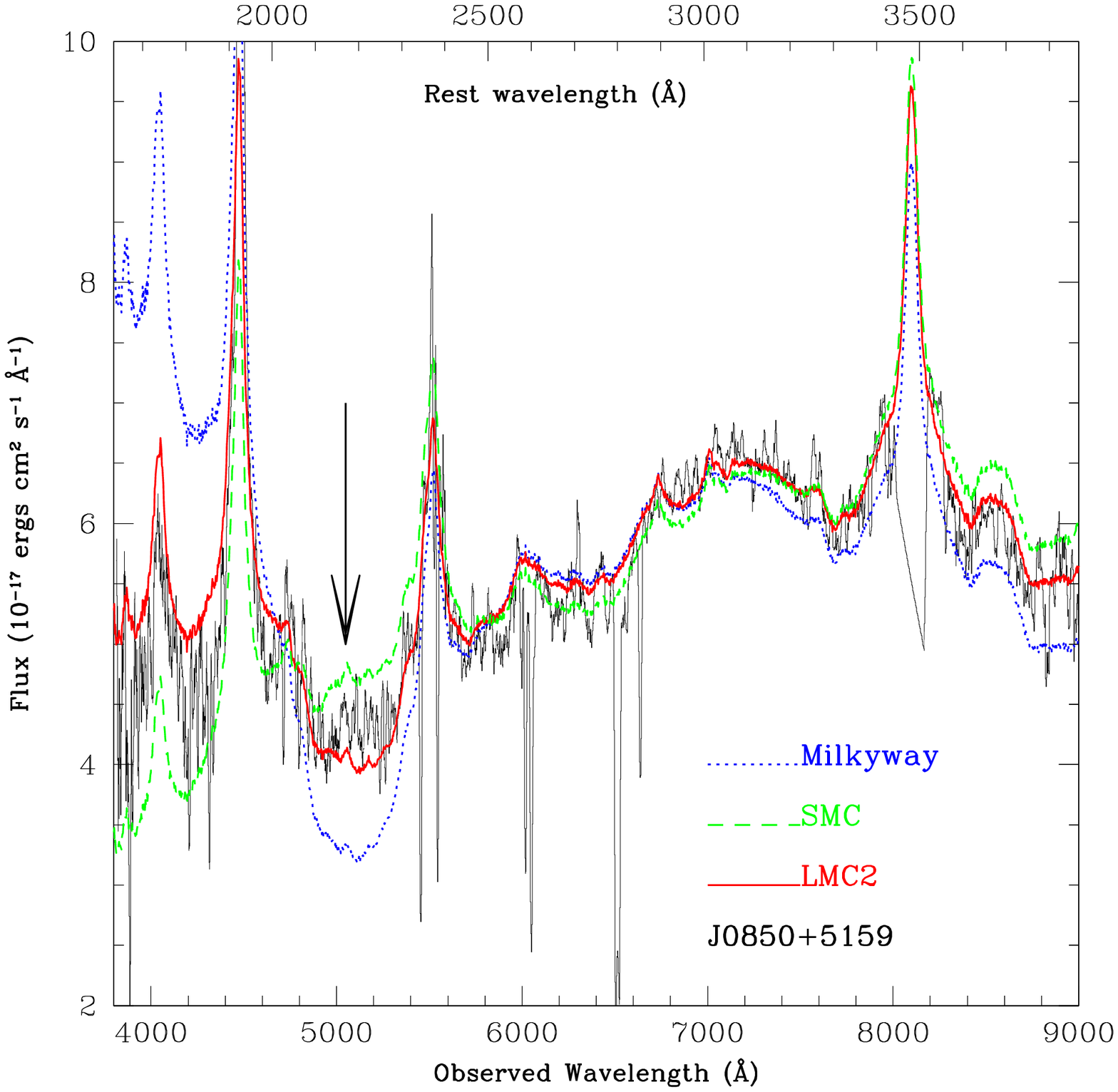,height=7.0cm,width=8.cm,angle=0}

\psfig{figure=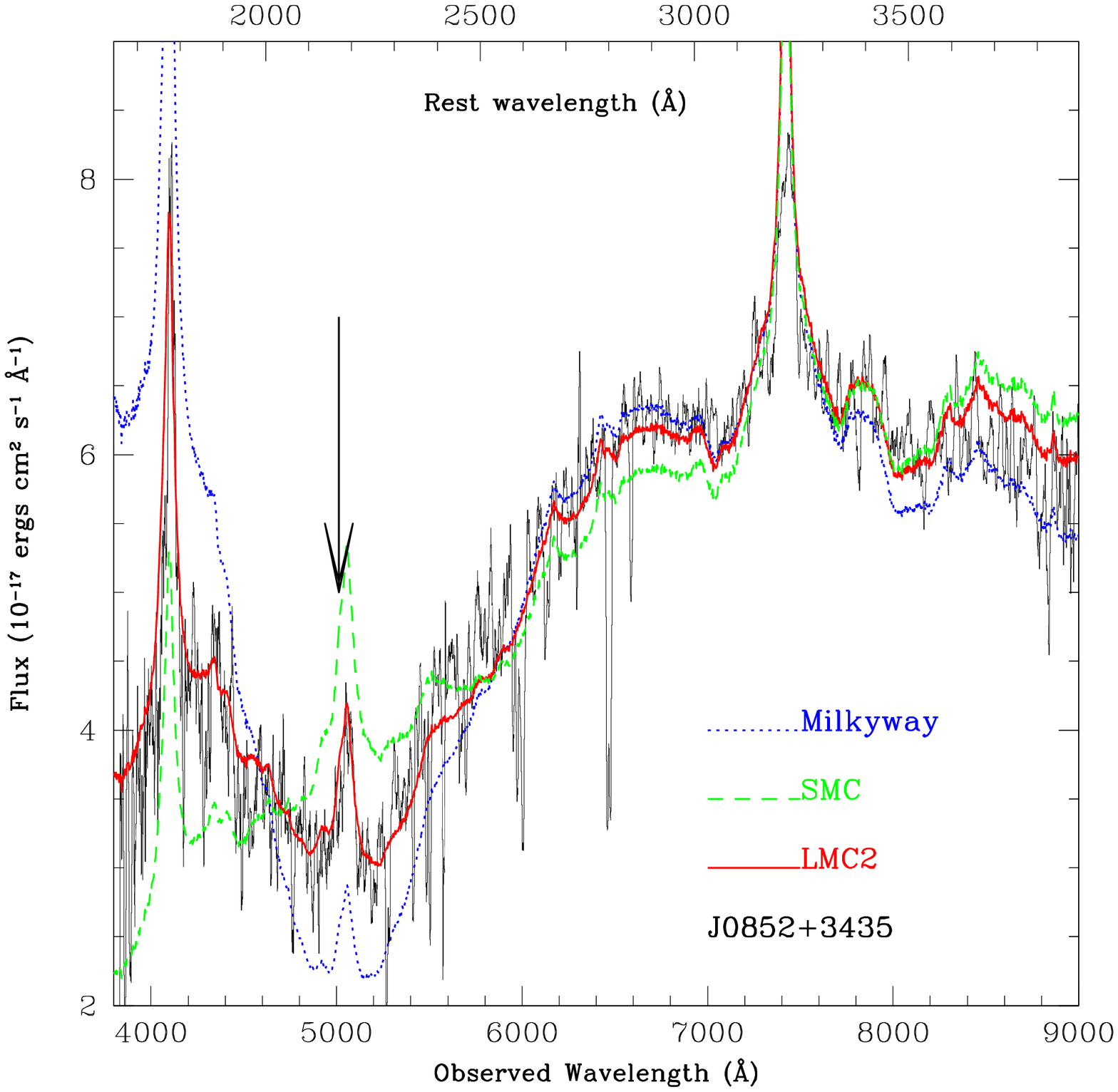,height=7.cm,width=8.cm,angle=0}
%}}
\caption[]{The SDSS spectrum of J0850+5159 (top) and
J0852+3435(bottom) are fitted with 
SDSS composite spectrum  corrected using average 
extinction curves from the Milky Way (dotted), the LMC2 Supershell (solid) and the SMC (dashed).
Rest wavelengths at \zabs are indicated at the top of the figure. The arrow marks the location of 2175 {\AA} feature at \zabs. For presentation purpose the observed spectrum
is boxcor smoothed by 10 pixels.
}
\label{fig1}
\end{figure}

We have developed a robust $\chi^2$ minimization code 
using the IDL routine MPFIT\footnote{details of MPFIT IDL routine can be found from http://cow.physics.wisc.edu/$\sim$craigm/idl/fitting.html} that uses 
Levenberg-Marquardt technique to get best fit values of $A_{V_{\rm a}}$
and normalization of the flux scale compared to the 
composite spectrum. We use LBQS, HST and SDSS QSO
composite spectra given in Francis et al. (1991), 
Zheng et al. (1997) and Vanden Berk et al. (2001)
respectively.
First we shift
the composite spectrum to the QSO emission  
redshift. 
For each extinction curve we predict the reddened QSO 
spectrum by multiplying the shifted composite spectrum 
by $\exp[-\tau(\lambda)]$. We mask the wavelength ranges 
of strong emission and absorption lines.
By varying the flux normalization and $A_{V_a}$ (or E(B-V))
we match the observed 
spectrum with our model reddened spectrum
(see Fig.~\ref{fig1}).
Results of our 
best fit models with different extinction curves 
and associated reduced $\chi^2$ are
summarised in Table~\ref{tchi}. The error in $A_{V_a}$
includes the effect of errors in the parameters of the
extinction curve.
For both the systems the lowest value of $\chi^2$ is 
obtained for the dust extinction models with the 
extinction curve of LMC2-Supershell irrespective of
our choice of composite spectrum.

\section{Discussion on Individual systems}

\subsection {The $z_{\rm abs} = 1.3265$ absorption system towards 
J085042.21$+$515911.7}

The spectrum of this QSO (with  u-K $\sim$ 4.8 mag) 
shows a curvature around 2175~\AA~ in 
the rest frame of the Mg~{\sc ii}
system (see  top panel of Fig.~\ref{fig1}). 
The Milky Way extinction curve that fits the 2175 {\AA} 
feature over predicts the QSO flux below 4500~\AA.
 This is also the case (not shown in Fig.~\ref{fig1}) when
we use the average LMC extinction curve. 
The SMC extinction curve
that fits the observed spectrum around 5500~{\AA}
under predicts flux below 4500~{\AA}. The LMC2 extinction curve that has a shallow UV-bump
and relatively high extinction at rest wavelength
$\lambda_r < 2100$ {\AA} compared to 
that of the Milky Way (and of the LMC) reproduces the data 
better. We also fitted the spectra of $\sim$ 200 non-BAL 
QSOs with \zem =1.892$\pm$0.005 using SMC extinction
curve  and 
%close to that
%of J0850+5159, we 
find the probability of the 2175~{\AA}
feature being 
produced by
QSO-to-QSO spectral variation to be $\le$0.03. For the range of composite spectra considered here,
the mean $A_{V_a}$ is 0.74$\pm$0.07 (and E(B$-$V) = 0.27). 
Our best fit model prediction J = 16.7, H = 15.8 and K = 14.7 mag,
with a typical uncertainty of 0.2 mag,
agrees well with the observed  values, J = 16.99$\pm$0.23, 
H = 15.72$\pm$0.16 and K = 15.27$\pm$0.15, from
Strutskie et al. (2006). The excess flux predicted in K band can be mainly attributed to the known host 
galaxy contribution ($\sim$0.3 mag) in the SDSS composite 
spectrum (See Vanden Berk et al. 2001) at $\lambda_r>$ 6000\AA.

From Table~2 of Gordon et al. (2003) we have,
\begin{equation}
N({\rm HI})\sim {A_{V_a}\over \kappa} (6.97\pm0.67)\times 10^{21} {\rm cm}^{-2}.
\end{equation}
Here $\kappa$ is the ratio of dust-to-gas ratio in the absorption system
to that in  LMC. 
We can use the equivalent widths of weak
transitions (Table~\ref{sdssuv}) to estimate the column densities
of species and derive the depletion factors onto dust-grains assuming 
Zn is not depleted:
$-$0.5,$-$0.9, $-$1.0, $-$0.9, and $-$1.0 for [Si/Zn], [Cr/Zn], [Fe/Zn], 
[Ti/Zn],  and [Mn/Zn] respectively. 
Note that Zn~{\sc ii}$\lambda2062$ is blended with
a Cr~{\sc ii} absorption but we can remove the contribution from Cr~{\sc ii} 
by using the average $N$(Cr~{\sc ii}) derived from unblended lines.
Depletion is higher than what is typically seen 
in high-$z$ absorption systems and are intermediate between what is 
seen in the Cold and Warm phases in the Galactic ISM (Welty et al. 
1999). Clearly more than 90\% of the refractory metals 
are in dust as in the LMC. If we assume $\kappa\sim 0.9$ then we
get $N$(H~{\sc i}) = (5.73$\pm$1.10)$\times10^{21}$ cm$^{-2}$. 

Our GMRT spectrum shows very strong absorption that can be modelled with two
Gaussian components (see Table~\ref{gmrtres} and upper panel in Fig.~\ref{gmrtdet}). The radio spectrum of this source is
flat and
VLBA maps at 5 and 15 GHz show that more than 90\% of the flux 
density is in the unresolved
core (Taylor et al. 2005). Using $f_{\rm c} =0.9$ in Eq.~1,
we derive 
$N$(H~{\sc i}) = 3.02$\pm0.84\times10^{19}$ $T_{\rm s}$ cm$^{-2}$. 
From the total $N$(H~{\sc i}) inferred from extinction we can estimate
the spin temperature, $T_{\rm s}\sim190^{+124}_{-69}$ K. Note that this value is 
an upper limit as the dust-to-gas ratio may be larger than 1.
All this is consistent with the system being associated with a
cold neutral medium with dust properties similar to that
seen in LMC2 supershell sample of Gordon et al. (2003).
\par\noindent

\subsection{The $z_{\rm abs} = 1.3095$ absorption system towards J085244.74$+$343540.5}

%\par\noindent
This QSO has u-K$\sim$5.6 mag and shows a curvature 
around 2175~\AA~in the rest frame of the Mg~{\sc ii}
system (see bottom panel of Fig.~\ref{fig1}) that is stronger than for the previous
quasar. 
The Mg~{\sc ii} system at \zabs = 1.3095
shows absorption lines that are also stronger than that of the previous system.
The SDSS spectrum of this quasar is also 
well reproduced when applying
a LMC2 like extinction curve to the QSO composite spectrum
(with $A_{V_a}$ = 1.00$\pm$0.09 and E(B$-$V) = 0.36).  
{
Using spectra of $\sim$ 330 non-BAL QSOs with 
\zem=1.655$\pm$0.005, we find the probability of 
the 2175~{\AA}
feature being 
produced by
QSO-to-QSO spectral variation to be $\le$0.001.
}
{Our best fit model prediction J = 16.5, H = 15.2 and K = 14.2 mag
agrees reasonably with the observed  values, J = 16.75$\pm$0.14, H = 15.57$\pm$0.11 and K = 15.10$\pm$0.12, considering
the possible uncertainties in the QSO composite spectrum
discussed above. }
We derive lower limits on the column densities 
using the weaker metal lines.
The absence of Cr~{\sc ii}$\lambda$2056 and Fe~{\sc ii}$\lambda$2249 features 
together with the high equivalent width of the Zn~{\sc ii}+Cr~{\sc ii} blend at 
$\lambda_r = 2062$~\AA~ is consistent with higher depletion factor
and higher metallicity in this system.
The detection of Ca~{\sc ii} absorption strongly
supports this conclusion. From the SED fit and assuming $\kappa = 1$ we 
derive ${\sl N}$(H~{\sc i}) =  6.97$\pm1.30\times10^{21}$ cm$^{-2}$. 
This may well be an upper limit as $\kappa$ could be larger than 1.

This system, despite having $N$(H~{\sc i})$\kappa$ similar 
to that in the \zabs = 1.3265 towards J0850+5159, has 
a factor of two lower total integrated $N$(H~{\sc i}) from
21-cm absorption. 
This source has a flat radio spectrum and is unresolved
in the VLA A-array 8.4 GHz image\footnote{from 
CLASS (Cosmic Lens All-Sky Survey) and JVAS (Jodrell/VLA Astrometric Survey) archive website\\
http://www.jb.man.ac.uk/research/gravlens/class/gmodlist.html} having a resolution of 0''.255$\times$0''.236.
However, high resolution (few mas
scale) VLBA map is not available for this source. Therefore,
while
$f_{\rm c}$ could be close to 1 its exact value is still
uncertain. We get a constraint $N$(H~{\sc i})$f_{\rm c}$/$T_{\rm s}$ = 
$1.31\pm0.26\times 10^{19}$
cm$^{-2}$ K$^{-1}$. By comparing the two $N$(H~{\sc i}) 
estimates
we derive $T_{\rm s}/f_{\rm c} = 536^{+234}_{-88}$ K.  
As $f_{\rm c}$ can be less than unity and $\kappa$ is probably larger
than one, the above value is an upper limit for the gas kinetic temperature. This 
is consistent with the gas being part of a cold neutral medium. 

\section{Discussion}

During our recently 
completed GMRT survey to search for 21-cm absorption from 
intermediate redshift (1.10$\le z\le 1.45$) Mg~{\sc ii} systems, we discovered two red QSOs with strong $z\sim 1.3$ Mg~{\sc ii} absorption 
systems along the line of sight. 
The observed spectral energy distribution of these QSOs are consistent with the QSO
spectrum being reddened by dust in the intervening Mg~{\sc ii} systems.
We fitted the SDSS spectrum using three QSO SDSS composite spectra 
reddened by different
extinction curves.
We find that the dust properties of the absorbers are similar to what is seen in the LMC2 supershell.
In particular, we detect the 2175~\AA~ UV extinction bump in both 
individual spectra.
The neutral hydrogen column densities, $N$(H~{\sc i}), inferred from
the extinction are consistent with the absorbing gas being of high
column density (i.e log $N$(H~{\sc i})$\sim$21.7).
We used weak metal absorption lines to
estimate the column densities of ions of different 
species.
Inferred metallicities 
are consistent with near solar values and depletion factors are similar
to what is measured in the Cold Neutral Medium of LMC.

These two quasars are rather weak
($\sim 50$ mJy) in radio emission
and are not ideally suited for 21-cm absorption observations.
We nonetheless detect 21-cm absorption from both systems
thanks to the favorable physical conditions in the 
absorbing gas.
These are the first detections of 21-cm absorption 
at high-$z$ towards such faint background sources. 
The inferred spin temperatures in these systems are 
consistent with that of a cold neutral medium gas.

The metal line equivalent widths  
and E(B-V) about $0.3$ measured  in the two systems 
discussed here are higher than those derived 
from the SDSS composite spectra of Mg~{\sc ii}
and Na~{\sc i} absorption systems
(see York et al. 2006a; Wild, Hewett \& Pettini 2006).
The E(B-V) values are also higher than
the median E(B-V) found for star forming galaxies 
at $z$ = 2$-$3  and a factor of 2 less than that
measured in submillimeter-selected galaxies
(see Figure 15 of Wild, Hewett \& Pettini 2006). 
The dust content and $N$(H~{\sc i}) in the two systems discussed in this
paper are comparable to that observed in the 21-cm system toward
AO 0234+164. 
Thus these two systems are ideally suited for high resolution
spectroscopic investigation of physical conditions in the
interstellar medium of the corresponding absorbing galaxies. 

\section{acknowledgements}

We thank the referee for very useful comments, 
Rajaram Nityananda for encouragement, and GMRT staff for 
their co-operation during the observations. 
The GMRT is an international facility run by NCRA-TIFR.
We acknowledge
the use of SDSS spectra from the archive (http://www.sdss.org/).
RS and PPJ gratefully 
acknowledge support from the Indo-French
Centre for the Promotion of Advanced Research.

%%%%%%%%%%%%%%%%%%%%%%%%%%%%%%%%%%%%%%%


\begin{thebibliography}{}

\bibitem{} Carilli, C. L., Menten, K. M., Reid, M. J.,
Rupen, M. P., Yun, M. S. 1998, ApJ, 494, 175

\bibitem{} Curran S. J., Whiting M. T., Murphy M. T., 
Webb J. K., Longmore S. N., Pihlstr\"om, Y. M.,
Athreya, R.,  Blake, C. 2006, MNRAS, 371, 431

\bibitem{} Ellison S., York B. A., Murphy M. T., Zych B. J., Smith A. M., Sarre P.
2008, MNRAS, 383, L30.

\bibitem{} Francis P., Hewett P. C., Foltz G., Chaffee F. H.,Weymann R. J., Morris S. L. 1991, ApJ, 373, 465
 
\bibitem{} Gordon, K. D., Clayton, C. G., Misselt, K. A., Landolt, A. U., Wolff, M. J. 2003, ApJ, 594, 279

\bibitem[]{} Gupta N., Salter C. J., Saikia, D. J., Ghosh, T., Jeyakumar, S. 2006, MNRAS, 356, 1509

\bibitem[]{} Gupta N., Srianand R., Petitjean P., Khare P., Saikia D. J., York D.
2007, ApJL, 654, L111

\bibitem[]{} Heng, K. et al., 2008, astrop-ph/0803.2879v2 

\bibitem[]{} Ishwara-Chandra C. H., Dwarakanath, K. S.,
Anantharamaiah K. R. 2003, JAA, 24 37.
 
\bibitem[]{} Junkkarinen, V. T., Cohen, R. D., Beaver, E. A., Burbidge, E. M., Lyons, R. W. Madejski, G. 2004, ApJ, 614, 658

\bibitem{} Kulkarni, V. P., York, D. G., Vladilo, G., Welty, D. 2007, ApJL, 663, 81

\bibitem{} Ledoux, C., Petitjean, P., Srianand, R. 2003,
MNRAS, 346, 209

\bibitem{} Misselt, K.A., Clayton, C. G., Gordon, K. D. 1999, ApJ, 515, 128

\bibitem{} Noll, S., Pierini, D., Pannella, M., Savaglio, S. 2007, A\&A, 472, 465

\bibitem{} Noterdaeme P., Ledoux C., Petitjean P., Srianand R. 2008, A\&A, 481, 327

\bibitem{} Pettini M., Smith L.J., Hunstead D. L., 1994, ApJ, 426, 79

\bibitem{} Petitjean P., Ledoux C., Noterdaeme P., Srianand R. 2006, A\&A, 456, L9

\bibitem{} Skrutskie, M. F., et al. 2006, AJ, 131, 1163

\bibitem{} Srianand, R., Petitjean, P., Ledoux, C.,
Ferland, G.,  Shaw, G., 2005, MNRAS, 362, 549

\bibitem{} Taylor, G. B. et al. 2005, ApJS, 159, 27

\bibitem{} Vanden Berk, D. et al. 2001, AJ, 122, 549
%
\bibitem{} Wang J., Hall P. B., Ge J., Li A., Schneider D. P. 2004, ApJ, 609, 589

\bibitem{} Welty D., Frisch P. C., Sonneborn G., York D. G. 1999, ApJ, 512, 636

\bibitem{} Wiklind T., Combes F. 2006, Natur, 379, 139
\bibitem{} Wild V., Hewett P., Pettini, 2006, MNRAS, 367, 211
%
\bibitem{} Wolfe A.M., Wills B. J. 1977, ApJ, 218, 39.
%
\bibitem{} Wolfire M. G., McKee C. F., Hollenbach D., Tielens A. G. G. M., Bakes E. L. O. 1995, ApJ, 443, 152.
%
\bibitem{} York B. A., Ellison S. L., Lawton B., Churchill C. W., Snow T. P., Johnson R. A.,
Ryan S. G. 2006, ApJL, 647, 29

\bibitem{} York D. G., et al. 2006a, MNRAS, 367, 945

\bibitem{} Zheng W., Kriss G. A., Telfer R. C., Grimes J. P., Davidsen A. F. 1997, ApJ, 475, 469.
\end{thebibliography}
\end{document}